\documentclass{aa}
\usepackage[usenames]{color}
\usepackage{graphicx}
\usepackage{amsmath}
\usepackage{amssymb}
\usepackage{xspace}
\usepackage{ulem}
\usepackage{amsmath}
\usepackage{natbib}
\usepackage{array}
\usepackage{rotating}
\usepackage{multirow}

\newcommand{\Msol}{\ensuremath{\text{M}_{\odot}}}

\usepackage{soul}

\begin{document}

\title{An analytic distribution function for a massless cored stellar system in a cuspy dark matter halo}
\titlerunning{Cored light profiles in cuspy dark halos}

\author{Maarten A. Breddels
  \and
  Amina Helmi
}
\institute{Kapteyn Astronomical Institute, University of Groningen, P.O. Box 800, 9700 AV Groningen, The Netherlands}
\authorrunning{M.A. Breddels \& A. Helmi}


\abstract{
We demonstrate the existence of distribution functions that
  can be used to represent spherical massless cored stellar systems embedded
  in cuspy dark matter halos with constant mildly tangential velocity
  anisotropy. In particular, we derive analytically the functional
  form of the distribution function for a Plummer stellar sphere in a
  Hernquist dark halo, for $\beta_0 = -0.5$ and for different degrees of
  embedding. This particular example satisfies the condition that the central
  logarithmic slope of the light profile $\gamma_0 > 2 \beta_0$. Our models have velocity dispersion profiles
  similar to those observed in nearby dwarf spheroidal galaxies. Hence they 
  can be used to generate initial conditions for a variety of
  problems, including N-body simulations that may represent dwarf galaxies in the Local Group.
}

\maketitle

\section{Introduction}

In the concordance $\Lambda$CDM cosmological model, galaxies are expected to be
embedded in massive dark matter halos. Recently, much focus has been
placed on measuring and modeling the internal dynamics of dwarf
spheroidal galaxies (dSph) as these systems have very high mass-to-light
ratios and appear to be dark matter dominated at all radii
\citep[see the recent reviews by][]{Walker2012,BHB2013}. Particular emphasis has been
put in establishing the characteristics of the host dark matter halos
and to determine whether their properties are consistent with those
expected in the context of the $\Lambda$CDM model \citep{Stoehr2002,
  Strigari2010}. More specifically, an open question is whether the dSph
satellites of the Milky Way could be embedded in density profiles
that are centrally cusped such as the NFW profile
\citep{NFW1996ApJ...462..563N}.

Much of this modeling work has been carried out using the Jeans
equations in the spherical limit \citep[e.g.][]{Lokas2001,Koch2007,Walker2007,walker2009uni,lokas2009,richardson2013}. 
The general goal has been to constrain the
dark matter content (i.e.\ to estimate the characteristic parameters
of given density profiles) by fitting the observed l.o.s.\ velocity
distributions, and more specifically the 2nd and 4th moments, i.e. the
dispersion and the kurtosis profiles. In Jeans modeling
the functional form of the velocity anisotropy needs to be
specified. A fundamental limitation of this approach is that the existence of a
distribution function, once a solution has been found, is not
guaranteed. Specifically, there is no assurance that a distribution
function that is positive everywhere (i.e. that it is physical), will exist.

Partly circumventing this problem, \citet{Wilkinson2002} introduced a family of parametric distribution functions that may be used
to represent spherical stellar systems with anisotropic velocity ellipsoids embedded in {\it cored} dark matter halos. More recently, the application of Schwarzschild's modeling
technique to dSph has allowed the consideration of more general density profiles (e.g. \citet{Jardel2012,Breddels2012arXiv,J2013,Breddels2013arXivb}). In this approach 
the distribution functions are obtained in a numerical fashion, and by construction 
they are positive everywhere.  However, these distribution functions have not been given in analytic form, and it may not even be plausible to find a simple
expression in the most general circumstances. 


The \citet{An2009} theorem provides an important constraint 
regarding distribution functions that may be associated to dSph. This theorem states that a system with a finite central
radial velocity dispersion must satisfy that the central value of the logarithmic slope of the stellar density profile $\gamma_0$
and the central velocity anisotropy $\beta_0$ be related through $\gamma_0 = 2 \beta_0$. This implies that if
the light profile is perfectly cored, as often assumed, i.e. $\gamma_0 = 0$, then
the velocity ellipsoid must be isotropic, {\it independently of the dark matter halo
profile} (which should be shallower than the singular isothermal sphere).
However, if the system is cold at the center, i.e. $\sigma_{r,0} = 0$, then the
only constraint is that $\gamma_0 > 2 \beta_0$, which in the case of cored stellar
profiles is satisfied by tangentially anisotropic ellipsoids \citep{Ciotti2010}. Since these
conditions refer to the intrinsic velocity dispersion, they do not impose strong
constraints on the line-of-sight velocity dispersion ($\sigma_{\rm los}$), which is the observable,
and one may obtain a flat $\sigma_{\rm los}$ profiles even if the system is intrinsically cold
at the centre.

Given the extensive modeling performed assuming cuspy dark matter halos and cored stellar profiles, 
the natural question that arises is whether physical distribution functions that can reproduce the properties of dSph exist in such cases. For example,  \citet{Evans2009} have shown that 
for a stellar Plummer profile with an isotropic velocity ellipsoid and a {\it strictly constant 
velocity dispersion profile}, the dark matter must follow a cored isothermal sphere. We
show here that this particular result cannot be generalized, and that (tracer) cored light
distributions can exist in equilibrium in cuspy dark matter halos, once the condition of constant velocity dispersion is relaxed.

In this letter, we present a distribution function that
represents a massless stellar system following a Plummer profile
embedded in a Hernquist dark matter halo, and which has a constant 
anisotropy $\beta = -1/2$. We focus on this particular example as
it is mathematically easy to manipulate, but also because it is
observationally sound. The surface brightness profiles of dSph are
well fit by Plummer models \citep{Irwin1995MNRAS.277.1354I}, and the velocity ellipsoids derived from
Schwarzschild models have radially constant, if slightly negative anisotropies \citep{Breddels2013arXivb}.
Therefore, models such as that presented below can be used, for example, to generate
initial conditions for an N-body simulation of a dwarf
galaxy resembling a dSph satellite of the Milky Way.

\section{Methods}
\label{sec:method}

\subsection{Generalities}
\label{sec:general-eq}

The distribution function of a spherical system in equilibrium can depend on energy $E$, and if the velocity
ellipsoid is anisotropic, also on angular momentum $L$: $f(E,L)$. It can be shown that when the
distribution function takes the form $f(E,L) = f_1(E) L^{-2\beta}$, with $\beta  = cst$, 
then this $\beta$ is the constant velocity anisotropy of the system. 

The functional form of the energy part of the distribution function
can be determined through an Abel equation, as outlined in Sec.~4.3.2
of \citet{BT}.  In that case (see their Eq.~4.67), we may derive $f_1(E)$
from
\begin{equation}
C_\beta \frac{d}{d\Psi} (r^{2\beta}\nu) = (\frac{1}{2} - \beta) \int_0^{\Psi} d \epsilon \frac{f_1(\epsilon)}{(\Psi - \epsilon)^{\beta + 1/2}}
\end{equation}
where $\nu (r)$ is the density, $\Psi(r) = -\Phi(r) + \Phi_0$
the relative gravitational potential, $\epsilon = -E = \Psi(r) -
1/2 v^2$ the relative energy, and $C_\beta$ is a constant. This equation
is valid for $-1/2 < \beta < 1/2$, and might be inverted using the
Abel integral to obtain an analytic expression for $f_1(\epsilon)$.
In the case of $\beta = -1$, an additional derivative is needed to
reach the Abel integral equation form, but the distribution function
may also be derived, now from
\begin{equation}
f_1(\epsilon) =  C'_{\beta = -1} \int_0^{\epsilon}\frac{d\Psi}{(\epsilon - \Psi)^{1/2}} \frac{d^3 (\nu/r^2)}{d \Psi^3}.
\label{eq:f1}
\end{equation}
These expressions are completely general, but only in the case
of gravitational potentials $\Psi(r)$  of simple mathematical form it is 
possible to invert and obtain $r$ as function of $\Psi$, and to easily 
compute all corresponding derivatives. 

The case of $\beta = -1/2$ is particularly simple and yields \citep[see Eq.~4.71 of][]{BT}
\begin{equation}
\label{eq:f1_E}
f_1(\epsilon) =  \frac{1}{2\pi^2} \frac{d^2 (\nu/r)}{d \Psi^2}\Big\rfloor_{\Psi = \epsilon}
\end{equation}

If the system were self-consistent, then the density and the potential
would be related through Poisson's equation \citep[see e.g.\
][]{Baes2002}. However, in the case of dSph, the gravitational potential is
largely determined by the dark matter, and the stars may simply be
considered as tracers. In this case, the above equations 
are still valid but the density is that of the stars $\nu_*(r)$, while we may assume the potential
to be that of the dark matter only. A priori, there is no guarantee that for example, the integral in Eq.~(\ref{eq:f1}) 
will converge, and that a physical solution, i.e.\ a positive distribution function leading to a stable system, will exist for
every combination of $\nu_*(r)$ and $\Psi(r)$.

\subsection{Plummer stellar sphere in a Hernquist dark halo, $\beta = -1/2$}
 
For mathematical convenience we assume 
that the gravitational potential is given by the Hernquist model \citep{Hernquist1990}, 
and as explained above, it is meant to describe the (dominant) contribution of the dark matter. 
Although this model is not cosmologically motivated, its density profile has the same
$r^{-1}$ limiting behavior in the inner regions as the NFW model. On the other hand, it has a finite
mass $M$, and a steeper fall off at large radii (as $r^{-4}$ instead of
$r^{-3}$). This is also why it is often used in the literature to set
up N-body simulations. The gravitational potential for the Hernquist model is
\begin{equation}
\label{eq:hq}
\Psi(r) = \Psi_0\frac{1}{1+r/b}, 
\end{equation}
where $b$ is the scale radius, and $\Psi_0 = G M/b$. For the stars we assume a Plummer profile
\begin{equation}
\label{eq:pl}
 \nu_*(r) = \frac{3}{4 \pi a^3} \frac{1}{(1 + r^2/a^2)^{5/2}},
\end{equation}
where $a$ the Plummer scale
length. Note that there are two characteristic lengthscales in the
problem, namely $a$ and $b$, and we will relate these using a
dimensionless parameter $\alpha = a/b$, and we expect that in general
$\alpha \le 1$ (i.e. that the stars are more concentrated than the dark matter halo in which they are embedded).

Using Eq.~(\ref{eq:hq}) we may thus express $r = r(\Psi)$ for the Hernquist
profile as
\begin{equation}
\label{eq:rhq}
r = b \frac{1 - \tilde{\Psi}}{\tilde{\Psi}}, \qquad \tilde{\Psi} = \Psi/\Psi_0.
\end{equation}
The energy part of the distribution function
$f_1(\epsilon)$ may now be computed explicitly from Eq.~(\ref{eq:f1_E}), using Eqs.~(\ref{eq:pl}) and (\ref{eq:rhq})
and after taking the corresponding derivatives, we find
\begin{equation}
f_1(\epsilon) = \frac{3}{8 \pi^3 (G M a)^2} \frac{\alpha^4 \tilde{\epsilon}^4}{(1 - \tilde{\epsilon})^3 (1 -2  \tilde{\epsilon} + (1 + \alpha^2) \tilde{\epsilon}^2)^{9/2}} p(\tilde{\epsilon})
\label{eq:f1_Efinal}
\end{equation}
where $\tilde{\epsilon} = \epsilon/\Psi_0$, $0 \le \tilde{\epsilon} \le 1$, 
and $p(\tilde{\epsilon})$ is the following polynomial
\begin{eqnarray}
p(\tilde{\epsilon}) = 30 - 108 \tilde{\epsilon}  + (132 - 5 \alpha^2) \tilde{\epsilon}^2 + 24 (-2 + \alpha^2) \tilde{\epsilon}^3 \nonumber \\
 - 3 (6 + 11 \alpha^2) \tilde{\epsilon}^4 +
 2 (6 + 7 \alpha^2 + \alpha^4) \tilde{\epsilon}^5.
\end{eqnarray}

\begin{figure}
\includegraphics[scale=0.4]{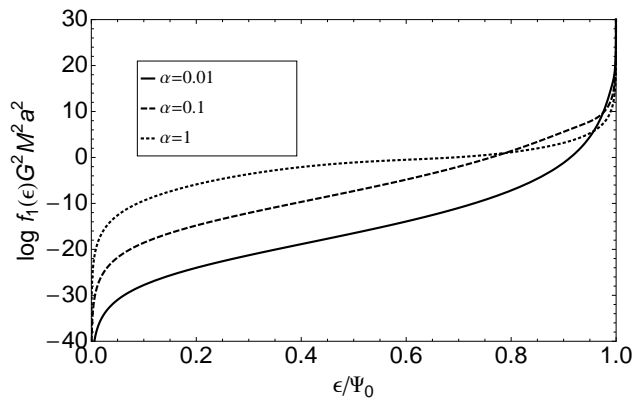}
\caption{Energy part of the distribution function for a Plummer
  stellar sphere of scale $a$ embedded in a Hernquist dark matter halo
  of mass $M$ and scale $b$, and with constant velocity anisotropy
  $\beta = -1/2$, as given by Eq.~(\ref{eq:f1_Efinal}). The various
  lines correspond to different degrees of embedding $\alpha = a/b$ of the stars in
  the (same) dark matter halo, $\alpha = 0.01$ (solid), $0.1$ (dashed) and 1 (dotted). }
\label{fig:df_e}
\end{figure}

Figure \ref{fig:df_e} shows the functional form of the distribution
function $f_1(\epsilon)$ for different values of $\alpha$, namely
$\alpha = 0.01, 0.1$ and 1, that is, for different degrees of embedding of
the stars in the dark matter halo.  The distribution function is
well-behaved, it is continuous and positive everywhere and has a
positive slope, indicating that it is stable to radial modes \citep[see Sec. 5.5 of][]{BT}.

\begin{figure*}
\includegraphics[scale=0.6]{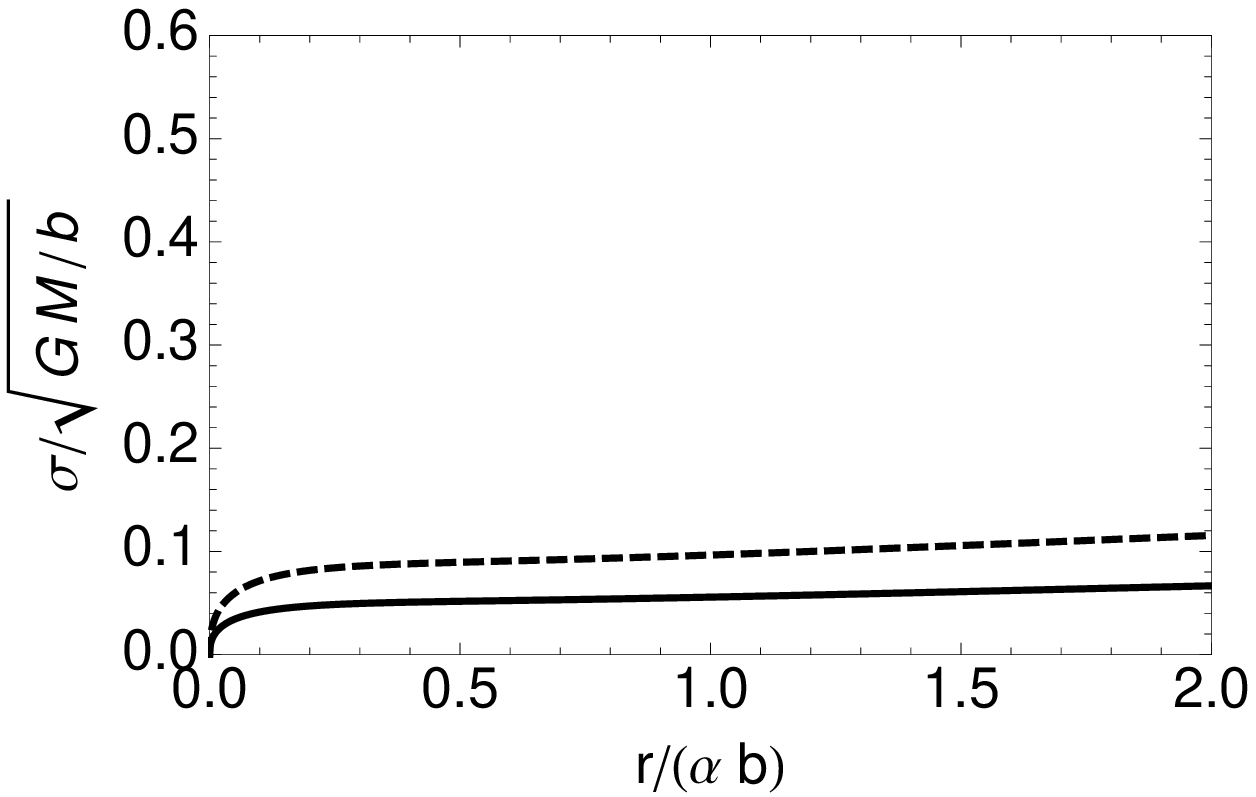}\includegraphics[scale=0.6]{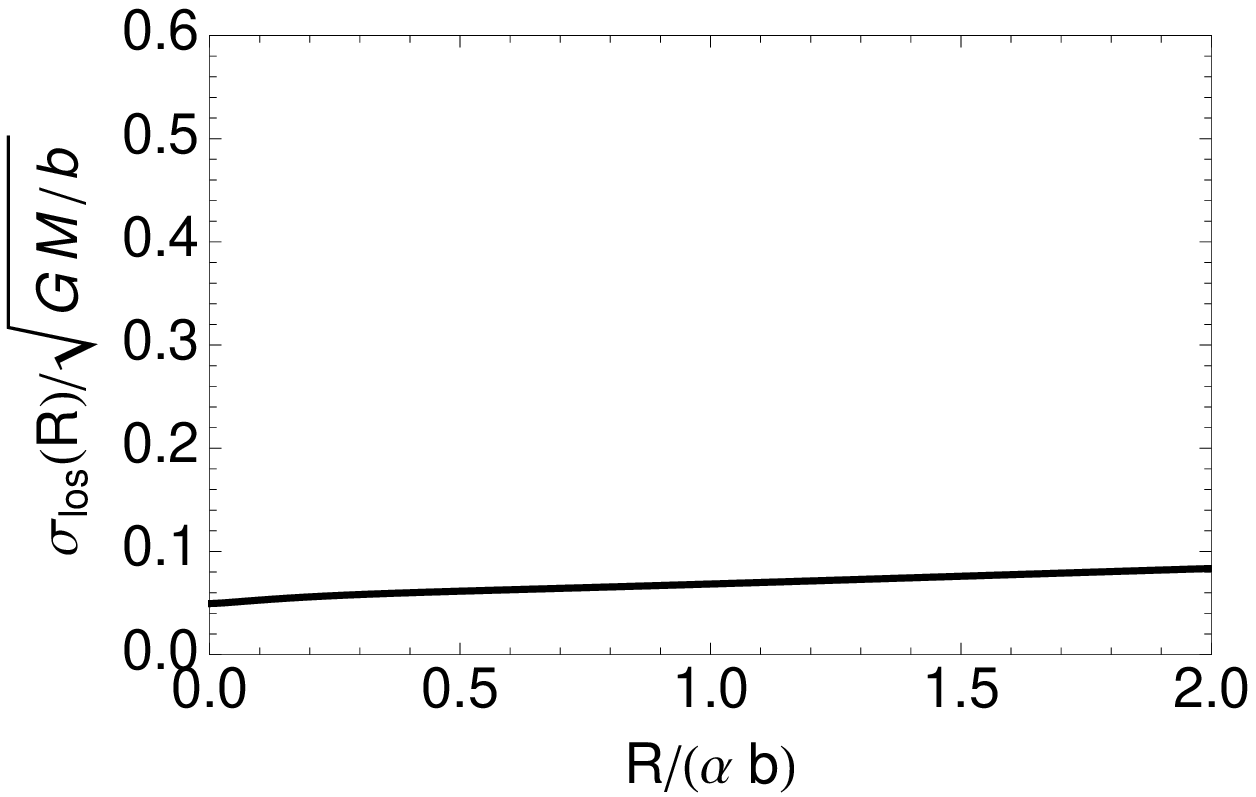}
\includegraphics[scale=0.6]{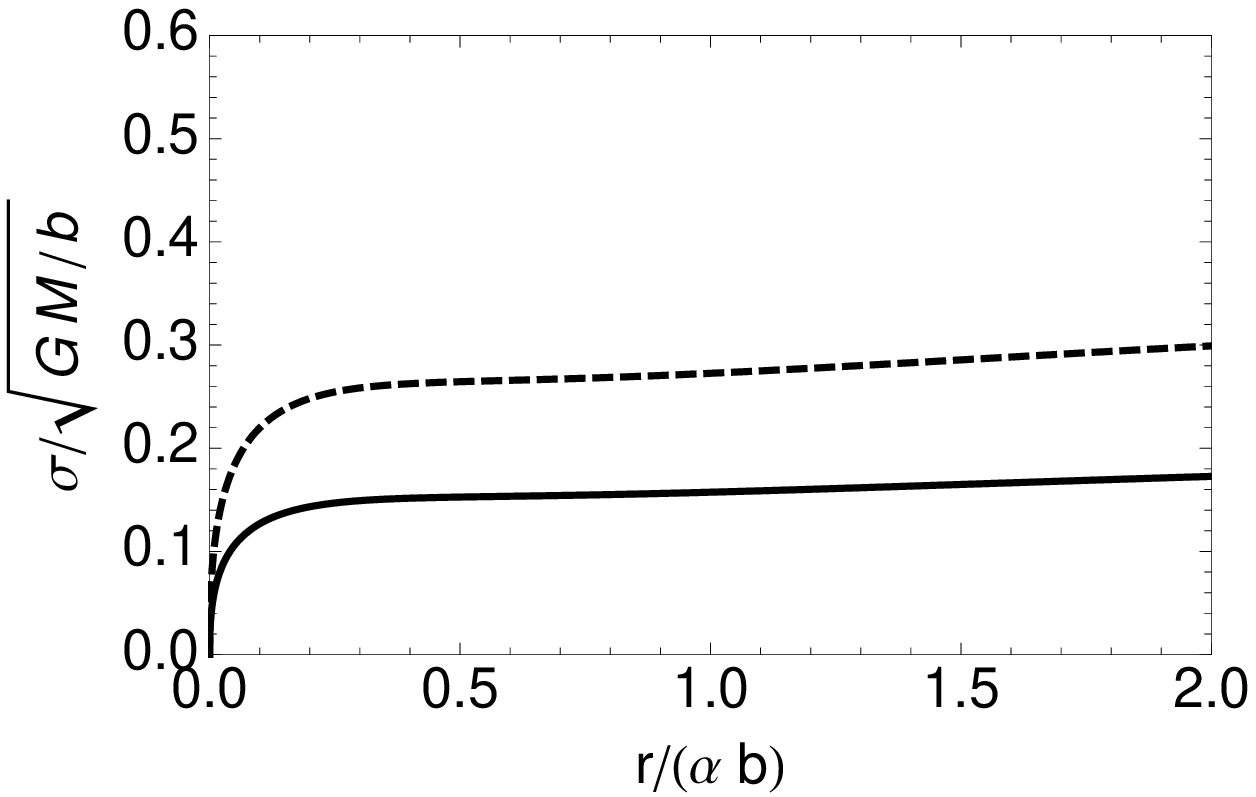}\includegraphics[scale=0.6]{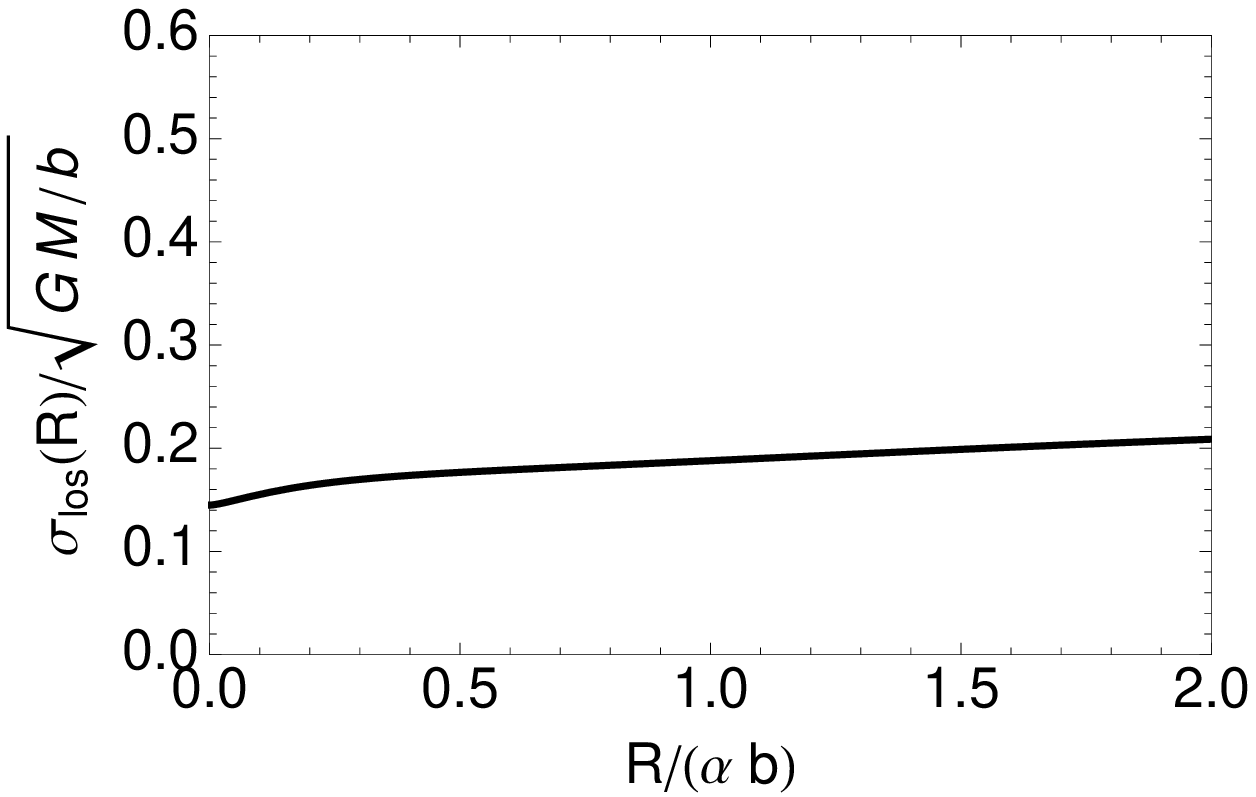}
\includegraphics[scale=0.6]{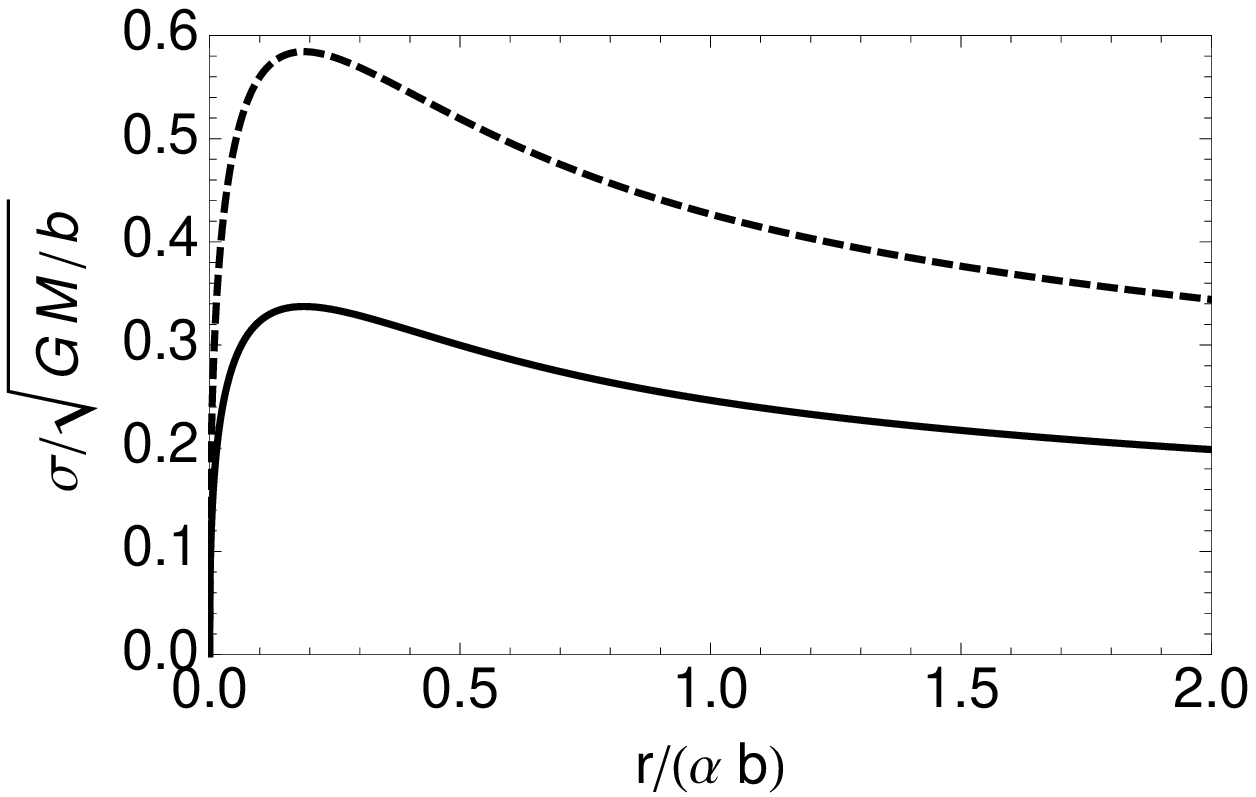}\includegraphics[scale=0.6]{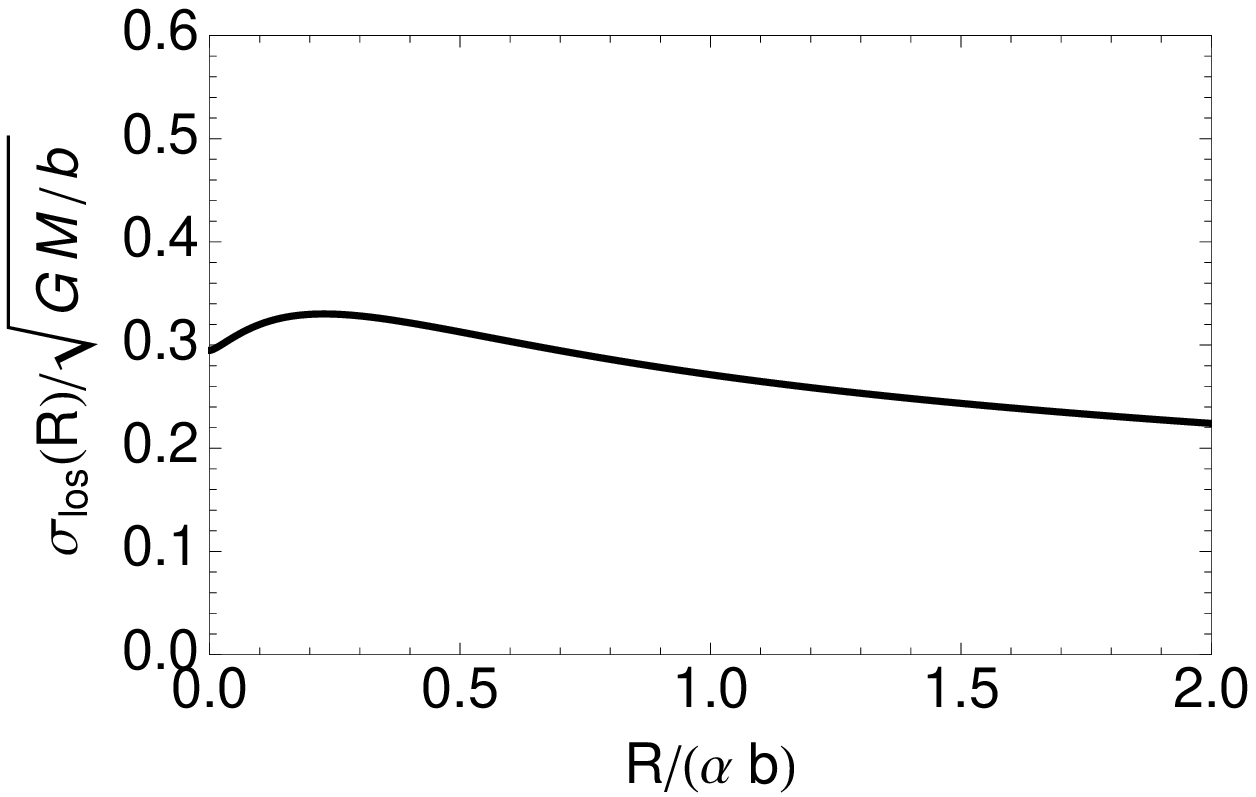}
\caption{Velocity dispersion profiles associated to the distribution functions presented in Fig.~\ref{fig:df_e} for different degrees of embedding, namely
$\alpha = 0.01, 0.1$ and 1 from top to bottom. {\bf Left column:} Intrinsic radial (solid) and tangential (dashed) velocity dispersions as function of
radial distance $r/a$. {\bf Right column:} Corresponding line-of-sight velocity dispersions as function of projected distance $R/a$. The resulting
$\sigma_{\rm los}$ profiles are relatively flat, especially for small $\alpha$, i.e. 0.01 and 0.1, and reach a finite
value at the centre. In general, the curves bare a good resemblance to the observed 
velocity dispersion profiles of stars in dSph satellites of the Milky Way, as determined in e.g.\ \citet{walker2009uni}.}
\label{fig:sigma}
\end{figure*}

We have checked that the density profile obtained by integrating this
distribution function over velocity space returns the Plummer
functional form.  The left column of Fig. \ref{fig:sigma} shows the velocity dispersion profiles
in the radial (solid) and tangential (dashed) directions for different values of
$\alpha$, and makes explicit the dependence of the internal kinematics on the
degree of embedding of the stars in the dark halo. For small $\alpha$ (top and middle left panels) 
the velocity dispersion profile is relatively flat for $r > 0.1 r/a$. Since the properties of
the halo are fixed by the mass $M$ and the scale $b$, we note that the
velocity dispersion has a smaller amplitude for smaller values of $\alpha$, as expected.


In the right column of Fig. \ref{fig:sigma} we have plotted the resulting l.o.s.\ velocity dispersion profiles for the different values of 
$\alpha$ explored. It shows that these profiles are relatively flat with projected radius over the range $0 < R/a < 2$,
which is similar to that probed by the observations the kinematics of stars in the dSph satellites of the Milky Way. If we set 
$M = 10^9 \Msol$ and $b = 2.5$ kpc, then the system with $\alpha = 0.01$ (top panels) would have
a velocity dispersion of $\sim 2$~km/s and $a \sim 25$ pc. On the other hand, if $\alpha = 0.1$ (middle panels), 
then $a = 250$ pc, and $\sigma_{\rm los} \sim 6.3$~km/s at the centre. This case could for example, represent
systems akin the Carina, Sextans or Ursa Minor dSph. A more massive halo would probably be required
if the aim is to represent systems like Sculptor or Fornax, as this would allow a better matching to the 
central value of $\sigma_{\rm los}$. 


\section{Conclusions}
\label{sec:end}

By example, we have demonstrated that a massless 
Plummer stellar system
embedded in a Hernquist dark matter halo constitutes a plausible
physical configuration. We have explicitly derived the form of the
distribution function for the case of a tangential anisotropy $\beta =
-1/2$ and for different degrees of embedding of the stars, as
quantified by the ratio of scalelengths parameter $\alpha = a/b$.  This
distribution function is positive for the values of $\alpha = 0.01 -
1$ and also leads to a system that is stable to radial modes, as
$\partial f/\partial \epsilon > 0$. The line-of-sight velocity
dispersion profiles characteristic of this family of distribution
functions resemble those observed for dSph, and hence can be used to represent these systems. They
satisfy the \citet{An2009} theorem, namely that $\gamma_0 > 2 \beta_0$, but clearly not the equality
condition. We have also explored the $\beta = -1$ case, and also found an analytic physical solution, though
this is more cumbersome mathematically and hence not presented here. 

\section*{Acknowledgments}
We are grateful to Tim de Zeeuw for encouragement and useful discussions,  and to J. An \& W. Evans for 
various interactions that led to the work presented here. We acknowledge financial support from 
NOVA (the Netherlands Research School for Astronomy), and 
European Research Council under ERC-StG grant GALACTICA-240271. 

\bibliography{refs}

\end{document}